\documentstyle[aps,prb,epsfig,twocolumn]{revtex}

\begin{document}

\bibliographystyle{prsty}

\draft

\wideabs{
\title{ {\bf Yes, 60 years later we are still working hard on vortices } }
\author{ {\bf Quantized Vortex Dynamics and Superfluid Turbulence  } }
\address{ Edited by C.F. Barenghi, R.J. Donnelly, and W.F. Vinen  \\
                Published by Springer, 2001.         \\
                Reviewed by Ping Ao (Oct. 2003)     }

\maketitle

}

To describe HeII, a bosonic quantum fluid, Lev Landau proposed a two-fluid theory in 1941 based on the concept of elementary or gapless excitations.  The success of this concept was later extended to fermionic quantum fluids. It has become a cornerstone in our understanding of physics in condensed states. Nowadays no question on it would be asked by any busy researcher. What less well-known is that Landau's 1941 paper also started the intriguing history on topological excitations, vortices:  Their existence was forbidden in his theory because Landau feared that it might destroy the rigidity of the superfluid state. His famous critical velocity criterion on supercurrent was not based on topology.  One of the consequences of this misconception was the delayed publication of the classical paper by A.A. Abrikosov, co-winner of 2003 Nobel Prize, on the existence of vortices and of vortex lattice in superconductors. Till these days the vortex dynamics has been remaining a fascinating research topic. R.J. Donnelly and W.F. Vinen, two of the three editors, are witnesses of and principal contributors in this long quest to understand dynamics of topological excitations.

The book is timely: The last major summary of the field was given by R.J. Donnelly in 1992, almost ten years ago; The discovery of high Tc superconductors has led to a tremendous progress in the statistical understanding of vortex states in superconductors during 1990's; and the study of topological excitations in Bose-Einstein condensates (BEC's) discovered in the middle 1990's begins to provide a fresh look on structure and dynamics of topological excitations.  Many world-renowned experts have contributed to it. I already mentioned Donnelly and Vinen. Here is an additional but by no means an exhaustive list: A.L. Fetter, P.V.E. McClintock, L.M. Pismen, E. Varoquaux.  

Let me give a random sampling of the book along with my comments. 
P.H. Roberts and N.G. Berloff provided a lucid presentation of vortex structure from a nonlinear Schrodinger equation (NLSE) with the normalization to superfluid density, though I am afraid that NLSE cannot be identified as the GP equation and it is latter which has not been derived by first principles in HeII.  Nevertheless, for anyone wishes to gain a quick understanding of vortex structure, their contribution is a must.  S. Rica further explored a very interesting consequence, the nucleation of vortices, within NLSE, a process vital in the understanding of the stability of supercurrent carrying states. The tools used in study of vortex dynamics in suerpfluids and superconductors start to be applied to the corresponding part in BEC's. The contribution by H.P Buechler, V.B. Geshkenbein, and G. Blatter is such an example. However, they did not check a `no-go theorem' in quantum field theory which implies that starting from their Eq.(3) the quantum nucleation is impossible. This throws a serious doubt on their quantum nucleation rate formula. 
R.J. Donnelly presented a concise summary of the experimental status on the superfluid turbulence. Superfluid appears to be a right system for such a study: The Reynolds number of superfluid is by definition infinite, hence obviously in the large Reynolds number regime desired for turbulence. It would be interesting to explore the connection between the superfluid turbulence and complicated vortex states in superconductors. The most interesting contribution in the book is by W.F. Vinen. With his forceful insight and motivated by an earlier study, Vinen stressed the importance of study of superfluid turbulence in a regime where normal fluid is negligible. It may reveal the importance of vortex-phonon interaction in this regime. As an emerging phenomenon, the superfluid turbulence may not depend on many microscopic details, particularly on those of mutual friction coefficients plaguing the interpretation of experimental data, in much the same way as that in the vortex states in superconductors. Vinen's suggestion should have an important methodological implication. It seems to be already supported by the numerical study such as presented in the contribution by of M. Tsubota, T. Araki, and S.K. Nemirovskii. It is most likely to expect a major progress in this direction.

Another line of approach to vortex dynamics starts from topology, which links naturally to the underlying quantum phase and microscopic theories. A long but partial list of major contributors is as follows:  F. London (1948), L. Onsager (1949), R.P. Feynman (1954), B.D. Josephson (1962), P.W. Anderson (1966). The history and current research activities from topological perspective were nicely summarized by D.J. Thouless (Topological Quantum Numbers in Nonrelativistic Physics, World Scientific, 1998).  
It is evident that vortex dynamics from topological point of view was missed in the reviewed book. Nevertheless I do take a notice that there is one contribution by L. Eaves discussing the connection to Quantum Hall effect and another by H.C. Chu and G.A. Williams discussing the generalization of Kosterlitz-Thouless transition, and that Onsager was referred in the contribution by V. Penna. 
Also conspicuously, through out the book there is no mentioning of the classical vortex dynamics experiment by Vinen in 1961 on a direct measurement of Magnus force, the transverse force on moving vortex. It is perhaps not surprising that though various corrections to the transverse forces are discussed in contributions by E. Sonin and by L.M. Pismen, no conditions were given on when and how their approximated results can be subjected to experimental tests.  
In view of more recent experimental indications of the absence of such corrections to the Magnus force, the experiment on Josephson-Anderson effect in superfluid of (R.E. Packard, 1998) and one on direct measurement of the transverse force in superconductors (X-M. Zhu, E. Brandstrom, B. Sundqvist, 1997), any knowledgeable reader would naturally ask Sonin and Pismen to provide testable conditions. 
Perhaps the field is still under the spell of Landau in its fondness towards classical hydrodynamic approach.

Apart from fields where it has already played an important role, such as superfluids, superconductors, BEC's, and Josephson junction arrays, one can easily find applications of vortex dynamics in many other fields of current interest, for example, the already mentioned quantum Hall effect, neutron stars, optical solitons, quantum computation, and, not the least, string theory.
As already suggested by the book, one would like to expect a new synthesis on vortex dynamics with insights from both classical fluid and topology based on more experiments from superfluid turbulence and BEC's. 
I believe the reviewed book will serve as a catalyst for a sequel to D.J. Donnelly's 1992 book, and recommend it to a prospective researcher as a good entry and a more experienced one for a quick survey, by keeping the need of topological consideration in mind.

\end{document}